\documentclass[aps,prd,preprint,superscriptaddress,showpacs,showkeys]{revtex4-1}
\usepackage{amsmath}
\usepackage{latexsym}
\usepackage{amsfonts}
\usepackage{graphicx}
\usepackage{hyperref}

\begin{document}
\title{The transient acceleration from time-dependent interacting dark energy models}
\affiliation{College of Mathematics and Physics, Chongqing University of Posts and Telecommunications,
Chongqing 400065, China}
\affiliation{MOE Key Laboratory of Fundamental Quantities Measurement, School of Physics, Huazhong
University of Science and Technology, Wuhan 430074, China}
\affiliation{State Key Laboratory of Theoretical Physics, Institute of Theoretical Physics,
Chinese Academy of Sciences, Beijing 100190, China}
\affiliation{Physics Division, National Technical University of Athens,
15780 Zografou Campus,  Athens, Greece}

\author{Xi-ming Chen}
\email{chenxm@cqupt.edu.cn}
\affiliation{College of Mathematics and Physics, Chongqing University of Posts and Telecommunications,
Chongqing 400065, China}
\affiliation{State Key Laboratory of Theoretical Physics, Institute of Theoretical Physics, Chinese Academy of Sciences, Beijing 100190, China}
\author{Yungui Gong}
\email{yggong@mail.hust.edu.cn}
\affiliation{MOE Key Laboratory of Fundamental Quantities Measurement, School of Physics, Huazhong
University of Science and Technology, Wuhan 430074, China}
\affiliation{State Key Laboratory of Theoretical Physics, Institute of Theoretical Physics, Chinese Academy of Sciences, Beijing 100190, China}
\author{Emmanuel N. Saridakis}
\email{msaridak@phys.uoa.gr}
\affiliation{Physics Division, National Technical University of Athens,
15780 Zografou Campus,  Athens, Greece}

\begin{abstract}
The transient acceleration which the current cosmic acceleration is not eternal is possible by introducing the interaction between dark matter and dark
energy. If the energy transfer is from dark energy to dark matter, then it is possible to realize the transient acceleration.
We study the possibility of transient acceleration by considering two time-dependent phenomenological interaction forms so that the energy
transfer increases as the universe evolves. Starting
from a simple and extending to a more complicated ansatz, we obtain
analytical expressions for the evolutions of the deceleration and the
various energy density parameters. We find the ranges of the parameters in the models for a transient acceleration.
\end{abstract}



\preprint{arXiv: 1111.6743}

\maketitle

\section{Introduction}

Recent cosmological observations support that the universe is experiencing
an accelerated expansion at late times \cite{hzsst98,scpsn98}. In principle there
are at least two ways to explain such a behavior, apart from the simple
consideration of a cosmological constant. The first approach is to modify the
gravitational sector itself, such as the Dvali-Gabadadze-Porrati model \cite{dgp},
f(R) gravity \cite{Carroll:2003wy,Starobinsky:2007hu,Hu:2007nk}, dGRT ghost-free massive gravity \cite{massgrav,Gong:2012yv},
etc. \cite{Capozziello:2003tk,Nojiri:2006ri}, obtaining a modified cosmological dynamics.
The other approach is to modify the content of the universe by introducing
dark energy which can be based on a canonical scalar field
(quintessence) \cite{peebles88,wetterich88,quintessence,track1,track2}, a phantom field
\cite{phantom,Caldwell:2003vq}, or the combination of quintessence and phantom fields in a unified
scenario \cite{Feng:2004ad,Cai:2009zp} (see Refs. \cite{copelandde,Padmanabhan:2007xy,limiaode} for a
review).

However, the dynamical nature of dark energy introduces a new cosmological
problem, namely why the energy densities of dark energy and dark
matter are nearly equal today although they scale independently during
the expansion history \cite{Liddle:1998xm,Guo:2006ab,Dutta:2009yb,Sahni:1998at,Uzan:1999ch,Bartolo:1999sq,Faraoni:2000wk,Saini:1999ba,Sen:2000zk,
Nojiri:2003vn,Onemli:2004mb,Nojiri:2005sx,Saridakis:2009pj,Dutta:2009dr,Guo:2004fq,Feng:2004ff,Zhao:2006mp,
Nojiri:2005pu,Setare:2008si}. The elaboration of this ``coincidence'' problem led
to the consideration of generalized versions of the above models
with the inclusion of a coupling between dark energy and dark matter. Thus,
various phenomenological forms of interacting dark energy models
\cite{Wetterich:1994bg,Amendola:1999,Billyard:2000bh,Farrar:2003uw,Wang:2005jx,Mimoso:2005bv,Lazkoz:2006pa,
Gonzalez:2006cj,Amendola:2006qi,Boehmer:2008,Saridakis:2007wx,gongplb09,gongjcap09,GarciaCompean:2007vh,Chimento:2011dw,
Chimento:2011pk,Szydlowski:2005kv,Feng:2007wn,Wei:2007zs,heJCAP08,31,bb,pt,AbdallaPLB09,AbdallaPLB09a,
Olivares:2007rt,Chen:2008ca,Bean:2008ac,Wei:2007ut,Quartin:2008px,secondref,Jamil:2009eb,Chimento:2009hj,
Guo:2004xx,Guo:2004vg,Nunes:2000ka,Mota:2004pa,Manera:2005ct,Nunes:2004wn,Clifton:2006vm,Clifton:2007tn,
Nojiri:2009pf,Curbelo:2005dh,Gonzalez:2007ht,Zhang:2012zz} have
been constructed in order to fulfill the observational requirements.

Due to the lack of information about dark energy and dark matter,
thus in the phenomenological models the interaction terms
in general were assumed to be proportional to the energy density and its derivative, and
to the Hubble parameter. However, such forms restrict the
time-dependence of the interaction term to a small and peculiar subclass of
possibilities, so a different approach was followed
\cite{Wang:2004cp,Alcaniz:2005dg,Costa:2007sq,Jesus:2008xi,Costa:2009mv,Costa:2009wv,Costa:2010bn,Baldi:2010vv}.
Instead of giving particular forms for the interaction term, the effect
of the interaction on the evolution of dark matter was explicitly shown
by the parameter $\epsilon(a)$ through the solution
$\rho_{dm}=\rho_{dm0}a^{-3+\epsilon(a)}$, with $a$ the scale factor.
Thus one can obtain interesting
cosmological behaviors by choosing the form of $\epsilon(a)$.
In the present work, we consider general time-dependent interactions.
It proves that even very simple forms can
alleviate the coincidence problem, and lend the cosmic acceleration a
transient character \cite{Russo:2004ym,Blais:2004vt,Bilic:2005sp,Srivastava:2006xq,Carvalho:2006fy,
Bento:2008yx,Fabris:2009mn,Guimaraes:2010mw}. It was shown that it is problematic to define a set of observable
quantities analogous to the S-matrix for string theory in an eternally accelerating universe
due to the existence of event horizon \cite{Hellerman:2001yi}. Therefore the existence of a
transient acceleration not only explains the current cosmic acceleration, but also
avoids the above mentioned problem for string theory. Motivated by
M theory, Albrecht and Skordis proposed a particular potential with both
exponential and polynomial forms \cite{Albrecht:1999rm} which leads to
transient acceleration \cite{Barrow:2000nc}. The compactification of M theory
with time-dependent hyperbolic internal space also provides a transient acceleration \cite{Gong:2007nh}.
On the other hand, it was found that the current cosmic acceleration is slowing down \cite{star,huang,Wu:2010mu,Cai:2011px},
suggesting that the current cosmic acceleration may not be eternal. Therefore,
it is necessary to study the model with a transient acceleration without the coincidence problem.

The plan of the work is as follows: In section \ref{model} we construct
the time-dependent interacting dark energy scenario, starting from a
simple interaction form (subsection \ref{simplest}), and extending the analysis to a
more general interaction form (subsection \ref{General}). Finally, section
\ref{Conclusions} is devoted to the summary of the results.

\section{Time-dependent interacting dark energy}
\label{model}

Let us now construct the time-dependent interacting dark energy scenario.
Throughout the work we consider a flat Friedmann-Robertson-Walker metric
of the form $
ds^{2}=-dt^{2}+a^{2}(t)d\bf{x}^2$.
The evolution equations for the energy densities of dark energy
and dark matter (considered as dust for simplicity) are
\begin{eqnarray}\label{eom1}
\dot{\rho}_{dm}+3H\rho_{dm}=Q,
\end{eqnarray}
\begin{eqnarray}\label{eom2}
\dot{\rho}_{de}+3H(\rho_{de}+p_{de})=-Q,
\end{eqnarray}
with $p_{de}$ the dark energy pressure, $Q$ the interaction term, $H\equiv
\dot{a}/a$ the Hubble parameter, and dot denoting differentiation with
respect to $t$. Therefore, $Q>0$ corresponds to the energy transfer from dark
energy to dark matter, while $Q<0$ corresponds to the energy transfer from dark matter
to dark energy. In general, we may consider a general interaction between scalar field and matter \cite{Koivisto:2012za}
\begin{equation}
\label{difconfint}
\mathcal{L}=\sqrt{-g}\left[-\frac{\mathcal{R}}{2\kappa^2}-\frac{1}{2}g^{\mu\nu}\partial_\mu\phi\partial_\nu\phi-V(\phi)\right]
-\sqrt{-\tilde{g}}\mathcal{L}_m(\psi,\tilde{g}_{\mu\nu}),
\end{equation}
where $\kappa^2=8\pi G$ and
$$\tilde{g}_{\mu\nu}=C(\phi)g_{\mu\nu}+D(\phi)\partial_\mu\phi\partial_\nu\phi.$$ Then the interaction $Q$ takes the form
$$Q=-\frac{C'-2D(3H\dot{\phi}+V'+C'\dot{\phi}^2/C)+D'\dot{\phi}^2}{2(C+D(\rho_m-\dot{\phi}^2))}\rho_m\dot{\phi}.$$
The above interaction is a generalization of the scalar tensor theory of gravity written in Einstein frame with $C=e^{2\beta\kappa\phi}$ and $D=0$.
Dependent on the choices of $C$ and $D$, we may derive desired interaction form $Q$.
Although the coupling between matter and gravity is non-minimal, the solar system constraints and other physical requirements
are satisfied due to the screening mechanism \cite{Koivisto:2012za}.
In terms of $Q$, we can introduce the effective equation of state $w_{eff}$ from the energy conservation
equations (\ref{eom1}) and (\ref{eom2}) as follows:
\begin{equation}
\label{weffeq}
w^{eff}_{de}=w_{de}+\frac{Q}{3H\rho_{de}},\quad w^{eff}_{dm}=-\frac{Q}{3H\rho_{dm}}.
\end{equation}
The system of dynamical equations closes by considering one of the
Friedmann equations:
\begin{equation}\label{FR1}
H^{2}=\frac{\kappa^{2}}{3}\Big(\rho_{de}+\rho_{dm}+\rho_b\Big),
\end{equation}
\begin{equation}\label{FR2}
\dot{H}=-\frac{\kappa^2}{2}\Big(\rho_{de}+p_{de}+\rho_{dm}+\rho_b\Big),
\end{equation}
where we have also included the dust baryon density (one can also
straightforwardly include the radiation too).

\subsection{Simplest model}
\label{simplest}

Lets us now determine the form of the interaction term $Q$. As we
mentioned in the introduction, we start from the interaction forms considered in the literature.
In particular, in the literature the
interaction forms were chosen as $Q=\alpha_0\dot{\rho}_{de}$ and
$Q=3\beta_0H\rho_{de}$ with constants $\alpha_0$ and $\beta_0$,
we generalize them to
\begin{equation}
\label{Qsimple}
Q=3\beta(a)H \rho_{de},
\end{equation}
with a simple power-law ansatz for $\beta(a)$, namely:
\begin{equation}
\label{cas32}
\beta(a)=\beta_0 a^\xi.
\end{equation}
The effective equation of state $w_{eff}$ become
\begin{equation}
\label{weff2.1}
w^{eff}_{de}=w_{de}+\beta_0 a^\xi,\quad w^{eff}_{dm}=-\beta_0 a^\xi\frac{\Omega_{de}}{\Omega_{dm}}.
\end{equation}
Therefore, $w^{eff}_{dm}$ becomes negative when $\beta_0>0$.

Substituting this interaction form (\ref{Qsimple}) into  Eq. (\ref{eom2}),
we obtain
 \begin{equation}
\label{rhophi2}
\rho_{de}=\rho_{de0}\, a^{-3(1+w_0)}\cdot
\exp{\left[\frac{3\beta_0(1-a^\xi)}{\xi}\right]},
\end{equation}
where the integration constant $\rho_{de0}$ is the value of dark energy  at present,
and for simplicity we have considered the dark energy
equation-of-state parameter $w\equiv p_{de}/\rho_{de}$ to be a constant $w_0$.
Substituting Eq. (\ref{rhophi2}) into Eq. (\ref{eom1}), we get
the dark matter energy density,
 \begin{equation}\label{rhom2}
\rho_{dm}=f(a)\rho_{dm0},
\end{equation}
where $\rho_{dm0}$ is the value of $\rho_{dm}$ at present,
\begin{equation}
\label{f1}
f(a)\equiv  \frac{1}{a^3}\left\{1-
\frac{\Omega_{de0}}{\Omega_{dm0}}\frac{3\beta_0 a^{-3w_0}
e^{\frac{3\beta_0}{\xi}}}{\xi}\times  \left[a^\xi
E_{\frac{3w_0}{\xi}}\left(\frac{3\beta_0 a^\xi}{\xi}\right)-a^{3w_0}
E_{\frac{3w_0}{\xi}}
\left(\frac{3\beta_0}{\xi}\right)\right]\right\},
\end{equation}
and $E_n(z)=\int_1^\infty t^{-n}e^{-xt}dt$ is the
usual exponential integral function.
Note however that Eq. (\ref{rhom2}) is an
analytic expression, while in Refs.
\cite{Wang:2004cp,Alcaniz:2005dg,Costa:2007sq,Jesus:2008xi,Baldi:2010vv} the corresponding expressions were left as
integrals and were elaborated numerically. To ensure $\rho_{dm}$ to be positive,
when $\beta_0<0$, we require $\xi < 3w_0$.  Obviously, in the case of
non-interaction, that is for $\beta_0=0$, Eq. (\ref{rhom2}) recovers the standard
result $\rho_{dm}=\rho_{dm0}/a^3$.
It is now easy to use the Friedmann equation (\ref{FR1}) to define
the dimensionless Hubble parameter, namely
\begin{equation}
\label{hubeq2}
E^2(z)\equiv \frac{H^{2}}{H^{2}_0}=\Omega_{b0}a^{-3}+\Omega_{dm0}f(a)
+\Omega_{de0}\,a^{-3(1+w_0)}\,e^{\frac{3\beta_0(1-a^\xi)}{\xi}},
\end{equation}
where $\Omega_{i}\equiv\kappa^2\rho_{i}/3H^2_{0}$, and $\Omega_{i0}\equiv\kappa^2\rho_{i0}/3H^2_{0}$
are the present values of the energy density parameters.
Therefore, from Eqs. (\ref{rhophi2}), (\ref{rhom2}) and (\ref{hubeq2})
we can straightforwardly obtain the evolutions of the density parameters
as
 \begin{eqnarray}\label{Omegab2}
\Omega_b(a)=\frac{a^{-3}}{a^{-3}+A f(a)+B
\,a^{-3(1+w_0)}\,
e^{\frac{3\beta_0(1-a^\xi)}{\xi}}},
\end{eqnarray}
\begin{eqnarray}
\label{Omegam2}
\Omega_{dm}(a)=\frac{f(a)}{A^{-1}a^{-3}+ f(a)+A^{-1}B
\,a^{-3(1+w_0)}\,
e^{\frac{3\beta_0(1-a^\xi)}{\xi}}},
\end{eqnarray}
\begin{eqnarray}\label{Omegaphi2}
\Omega_{de}(a)=\frac{\,a^{-3(1+w_0)}\,
e^{\frac{3\beta_0(1-a^\xi)}{\xi}}}{B^{-1}a^{-3}+AB^{-1} f(a)+
\,a^{-3(1+w_0)}\,
e^{\frac{3\beta_0(1-a^\xi)}{\xi}}},
\end{eqnarray}
where
$A=\Omega_{dm0}/\Omega_{b0}$ and $B=\Omega_{de0}/\Omega_{b0}$.
The deceleration parameter
\begin{eqnarray}
\label{deceleration00}
q\equiv-\frac{\ddot{a}}{aH^2}=-1+\frac{3}{2}\left[\frac{
\Omega_b+\Omega_m+(1+w_0)\Omega _{de}} { \Omega_b+\Omega_m+\Omega
_{de}}\right],
\end{eqnarray}
is found to be
\begin{equation}\label{deceleration2}
q=-1+\frac{3}{2}\left[{\frac{a^{-3}+A
f(a)+B(1+w_0)\,a^{-3(1+w_0)}\,
e^{\frac{3\beta_0(1-a^\xi)}{\xi}}}{a^{-3}+A
f(a)+B\,a^{-3(1+w_0)}\,
e^{\frac{3\beta_0(1-a^\xi)}{\xi}}}}\right].
\end{equation}
If $\beta_0>0$ and $\xi>0$, then $a^{-3(1+w_0)}e^{3\beta_0(1-a^\xi)/\xi}\rightarrow 0$
and $q\rightarrow 1/2>0$ when $a\rightarrow \infty$, so the transient acceleration happens.

For the special case $\xi=0$ and $\beta_0\neq -w_0$,
the energy densities of the dark sectors are
\begin{equation}
\label{dedens1}
\rho_{de}=\rho_{de0}a^{-3(1+w_0+\beta_0)},
\end{equation}
\begin{equation}
\label{dmdens1}
\rho_{dm}=\rho_{dm0}a^{-3}\left[1+\frac{\Omega_{de0}}{\Omega_{dm0}}\frac{\beta_0}{w_0+\beta_0}\left(1-a^{-3(w_0+\beta_0)}\right)
\right].
\end{equation}
To guarantee $\rho_{dm}$ to be positive, we require $\beta_0\ge 0$ when $a\rightarrow \infty$ and $\beta_0<-w_0\Omega_{dm0}/(\Omega_{dm0}+\Omega_{de0})$ when $a\rightarrow 0$.
Therefore, the coupling constant $\beta_0$ must satisfy the condition $0\le \beta_0<-w_0\Omega_{dm0}/(\Omega_{dm0}+\Omega_{de0})$.
Using Eqs. (\ref{dedens1}) and (\ref{dmdens1}), we get
\begin{equation}
\label{decqeq1}
q=\frac{1}{2}+\frac{3}{2}w_0\Omega_{de0}\left[\frac{w_0\Omega_{de0}}{w_0+\beta_0}+
\left(1-\frac{w_0\Omega_{de0}}{w_0+\beta_0}\right)a^{3(w_0+\beta_0)}\right]^{-1}.
\end{equation}
At present, $q(a=1)<0$, so we require $w_0\Omega_{de0}<-1/3$.
To get transient acceleration, we require $q>0$ when $a\gg 1$, so we need $\beta_0>-w_0-1/3$ and $\beta_0\neq -w_0$.
Since current acceleration requires that $w_0<-1/(3\Omega_{de0})<-1/3$, so
$\beta_0>-w_0-1/3>0$. Combining with the physical condition that $\rho_{dm}$ is positive, we
find that $-w_0\Omega_{dm0}/(\Omega_{dm0}+\Omega_{de0})>\beta_0>-w_0-1/3$.

For the most special case
$\xi=0$ and $\beta_0=-w_0$, we get $\rho_{dm}=\rho_{dm0}a^{-3}[1+3\beta_0(\Omega_{de0}/\Omega_{dm0})\ln a]$.
If $\beta_0\neq 0$, $\rho_{dm}$ is negative either when $a\ll 1$ or $a\gg 1$, so we do not consider
the case $\xi=0$ and $\beta_0=-w_0$.

The cosmic acceleration is a transient one when the parameters satisfy the conditions:
(1) $-w_0\Omega_{dm0}/(\Omega_{dm0}+\Omega_{de0})>\beta_0>-w_0-1/3$ if $\xi=0$; (2) $\beta_0>0$ if $\xi>0$.
Up to now we derived analytical expressions for the
evolutions of the various density parameters, and the
deceleration parameter, with only the present values of the density parameter
and the equation of state parameter for dark energy as
free parameters. It is therefore straightforward to construct their
evolution graphs, using the observational values $\Omega_{de0}=
0.72$, $\Omega_{dm0}= 0.24$, $\Omega_{b0}= 0.04$, and $w_0=-0.98$ for quintessence and $w_0=-1.02$ for phantom
\cite{Bennett:2003bz,Tegmark:2003ud,Allen:2004cd,wmap7,gong12b}.
To have a clear picture of the above results, we take some particular values of $\xi$ and $\beta_0=\pm 0.01$ \cite{chenxm11}
as examples to plot the evolutions of dark energy and dark matter.

\begin{figure*}[htp]
\begin{center}
\includegraphics[width=0.8 \textwidth]{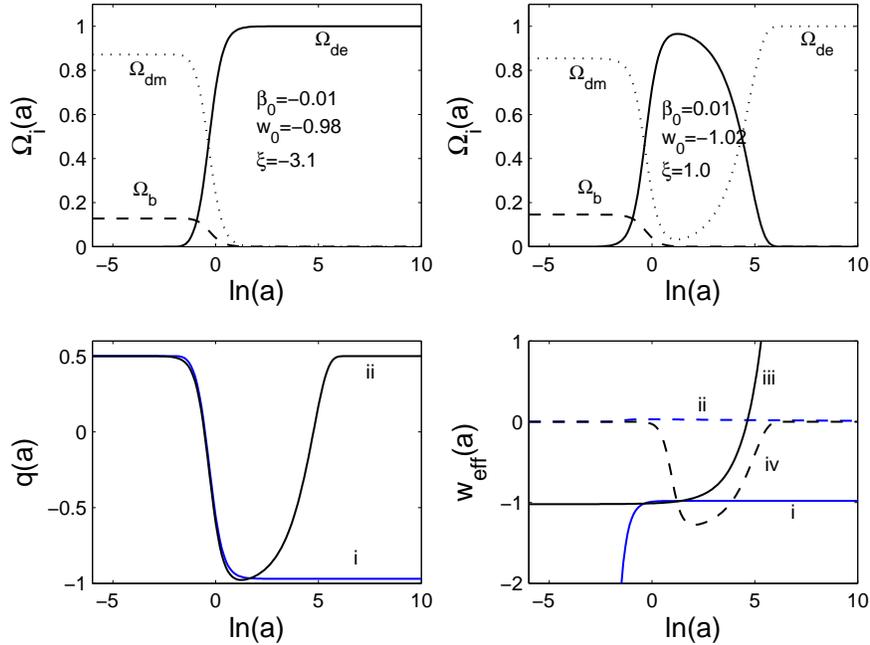}
\end{center}
\caption{The results for the simplest interacting model $Q=3\beta_0 a^\xi H \rho_{de}$.
Upper left panel: The evolutions of the various density parameters
for $\beta_0=-0.01$, $\xi=-3.1$ and $w_0=-0.98$.
Upper right panel: The evolutions of the various density parameters
for $\beta_0=0.01$, $\xi=1.0$ and $w_0=-1.02$.
Lower left panel: The corresponding
evolutions of the deceleration parameter $q$. Line (i) is for the parameters
$\beta_0=-0.01$, $\xi=-3.1$ and $w_0=-0.98$ and line (ii) is for the parameters
$\beta_0=0.01$, $\xi=1.0$ and $w_0=-1.02$. Lower right panel: the evolutions
of the effective equation of state parameters for dark energy (lines (i) and (iii)) and dark matter (lines (ii) and (iv)).
Lines (i) and (ii) are for the parameters
$\beta_0=-0.01$, $\xi=-3.1$ and $w_0=-0.98$, and lines (iii) and (iv) are for the parameters
$\beta_0=0.01$, $\xi=1.0$ and $w_0=-1.02$.
}
\label{mod2}
\end{figure*}

In the upper left panel of Fig.~\ref{mod2} we plot the evolutions of the
various density parameters with $\beta_0=-0.01$, $w_0=-0.98$ and $\xi=-3.1$,
corresponding to the energy transfer from dark matter to dark energy.

Due to the energy transfer from dark matter to dark energy, despite the
fact that the energy transfer decreases as time passes by ($\xi$ is negative),  we obtain the expected
result of complete dark energy domination in the future.
In the lower left panel of Fig.~\ref{mod2} we depict the
corresponding evolution of the deceleration parameter (line (i)). Clearly, we can see
that in this scenario, the late-time cosmic acceleration is permanent.

In the upper right panel of Fig.~\ref{mod2} we present the evolutions of the various density
parameters with $\beta_0=0.01$, $w_0=-1.02$ and $\xi=1.02$. It is clear that the
cosmic acceleration is transient. Because positive $\beta_0$ corresponds to the energy
transfer from dark energy to dark matter and positive $\xi$ means increasing energy transfer
as the universe evolves, so dark matter will finally become the dominant component.

In the phantom case, $w_0<-1$, we find that the interaction can not only save the
universe from a Big Rip \cite{GonzalezDiaz:2003bc,Kallosh:2003bq,Setare:2008pc,Capozziello:2009hc},
but also lead to a dark matter domination.
Additionally, in the lower left panel of Fig.~\ref{mod2} we
plot the evolution of the deceleration parameter (line (ii)). From the plot we can
clearly see that the present acceleration of the universe is transient
when both $\beta_0$ and $\xi$ are positive. This is a very interesting result
from the phenomenological point of view, and one of the main results of the
present work. The result of transient acceleration is quite general for interacting models with more and more energy transfer from dark energy to dark matter.

In the lower right panel of Fig.~\ref{mod2}, we show the evolutions of the effective equation of state parameters $w_{eff}$
for both dark energy and dark matter. We see that $w^{eff}_{de}$ becomes positive (line (iii))
and $w^{eff}_{dm}$ decreases to be negative first and increases to zero again  in
the future (line (iv)) due to the energy transfer from dark energy to dark matter when transient acceleration happens.
In the dark energy domination case, $w^{eff}_{dm}$ keeps to be non-negative and $w^{eff}_{de}$ keeps to be negative.

\subsection{General scenarios}
\label{General}

In this subsection we extend the previous analysis to more general
time-dependent interacting scenarios. In particular, we add $\alpha(a)\dot\rho_{de}$ and consider
\begin{equation}
\label{Qfull}
Q=3\beta(a)H \rho_{de}+\alpha(a)\dot{\rho}_{de},
\end{equation}
with a simple power-law ansatz for $\alpha(a)$:
\begin{equation}
\label{casefull}
\alpha(a)=\alpha_0 a^\eta,
\end{equation}
and the same power-law form (\ref{cas32}) for  $\beta(a)$.

Substituting this interaction form into Eq. (\ref{eom2}),
we obtain
\begin{eqnarray}
\label{f1full1}
  \rho_{de}=&\rho_{de0}
 a^{-3(1+w_0)} \left(\frac{1+\alpha_0
a^\eta}{1+\alpha_0}\right)^{\frac{3(1+w_0)}{\eta}}
\exp\left\{3\beta_0\xi^{-1}\times \right. \nonumber\\
&\left.\left[
\,_2F_1\left(1,\frac{\xi}{\eta};\frac{\eta+\xi}{\eta}
;-\alpha_0\right)-
a^{\xi}\,_2F_1\left(1,\frac{\xi}{\eta};\frac{\eta+\xi}{\eta}
;-a^\eta\alpha_0\right)\right]\right\},
\end{eqnarray}
where $\,_2F_1(a,b;c;z)$ is the
usual hypergeometric function, and
again we consider the dark energy
equation-of-state parameter to be a constant $w_0$. Note that this
expression coincides with Eq. (\ref{rhophi2}) when $\alpha_0=0$, as
expected.

Unfortunately, inserting the above solution (\ref{f1full1})
into Eqs. (\ref{Qfull}) and (\ref{eom1}), does not lead to an analytical expression for
$\rho_{dm}$. Since in the present paper we desire to provide analytical
results, so we restrict ourselves in the simpler scenario
 \begin{equation}
\label{case1}
\alpha(a)= \beta(a)= \alpha_{0} a^{\xi}.
\end{equation}
In this case,
Eq. (\ref{f1full1}) becomes
\begin{equation}
\label{rhodefull2}
\rho_{de}=\rho_{de0}a^{-3(1+w_0)}
\left(\frac{1+\alpha_0 a^\xi}{1+\alpha_0}\right)^{\frac{3w_0}{\xi}}.
\end{equation}
To ensure $\rho_{de}$ to be always real, we require $\alpha_0>0$.
The effective equation of state is
\begin{equation}
\label{weff2.2}
w^{eff}_{de}=w_{0}-w_0\alpha_0\frac{a^\xi}{1+\alpha_0 a^\xi},\quad
w^{eff}_{dm}=w_0\alpha_0\frac{a^\xi}{1+\alpha_0 a^\xi}\frac{\Omega_{de}}{\Omega_{dm}}.
\end{equation}
So $w^{eff}_{dm}$ is always negative. The solution for dark matter  is
 \begin{equation}\label{rhom2full}
\rho_{dm}=g(a)\rho_{dm0},
\end{equation}
where
\begin{eqnarray}
\label{f1full}
g(a)\equiv &\frac{1}{a^3}\left\{1-
\frac{\Omega_{de0}}{\Omega_{dm0}}
\left(\frac{3\alpha_0w_0}{3w_0-\xi}\right)
\left[\,_2F_1\left(1,1;2-\frac{3w_0}{\xi}
;-\alpha_0\right)-\right. \right.\nonumber\\
 &\left.\left. a^{\xi-3w_0}\left(\frac{1+\alpha_0
a^\xi}{1+\alpha_0}\right)^{\frac{3w_0}{\xi}}\,_2F_1\left(1,1;2-\frac{
3w_0}{\xi}
;-\alpha_0 a^\xi\right) \right] \right\}.\nonumber
\end{eqnarray}
 Note that in the case of
no-interaction, that is for $\alpha_0=0$, Eq. (\ref{rhom2full}) gives the
standard result $\rho_{dm}=\rho_{dm0}/a^3$. In fact, substituting Eq. (\ref{case1})
into Eq. (\ref{Qfull}), the interaction form becomes $Q=-3\alpha(a)w_0H\rho_{de}/(1+\alpha(a))$.
So the form of interaction is the same as Eq. (\ref{Qsimple}) except that now $\beta(a)=-\alpha(a)w_0/(1+\alpha(a))$.
The general scenario is just a special case of the the simplest scenario with $\beta(a)=-\alpha(a)w_0/(1+\alpha(a))$.
For $\xi>0$, $\beta(a)\rightarrow -w_0$ when $a\rightarrow \infty$
and $\beta(a)\rightarrow 0$ when $a\rightarrow 0$ if $\alpha_0\neq 0$.
It becomes the special case $\xi=0$ of the simplest model (\ref{cas32}) discussed in section \ref{simplest}.
Since $\beta(a)=-w_0>-w_0-1/3$ when $a\rightarrow \infty$,
so the acceleration is a transient one  for $\xi>0$ and $\alpha_0>0$. We may worry about the positivity of
$\rho_{dm}$ when $a\rightarrow 0$, in fact there is no problem because $\beta(a)\rightarrow 0$ when $a\rightarrow 0$.
For the case $\xi<0$,
$\beta(a)\rightarrow -w_0 \alpha_0 a^\xi \rightarrow 0<-w_0-1/3$ when $a\rightarrow \infty$,
so the acceleration is permanent.

The dimensionless Hubble parameter reads:
\begin{equation}
\label{hubeq2full}
E^2(z)\equiv
\frac{H^{2}}{H^{2}_0}=\Omega_{b0}a^{-3}+\Omega_{dm0}g(a)+
\Omega_{de0}\,a^{-3(1+w_0)}
\left(\frac{1+\alpha_0 a^\xi}{1+\alpha_0}\right)^{\frac{3w_0}{\xi}}.
\end{equation}
Thus, Eqs. (\ref{rhodefull2}), (\ref{rhom2full}) and
(\ref{hubeq2full}) give
 \begin{eqnarray}
 \label{Omegab2full}
\Omega_b(a)=\frac{a^{-3}}{a^{-3}+A g(a)+B
a^{-3(1+w_0)}
\left(\frac{1+\alpha_0 a^\xi}{1+\alpha_0}\right)^{\frac{3w_0}{\xi}}},
\end{eqnarray}
\begin{eqnarray}
\label{Omegam2full}
\Omega_{dm}(a)=\frac{g(a)}{A^{-1}a^{-3}+ g(a)+A^{-1}B a^{-3(1+w_0)}
\left(\frac{1+\alpha_0 a^\xi}{1+\alpha_0}\right)^{\frac{3w_0}{\xi}}},
\end{eqnarray}
\begin{eqnarray}
\label{Omegaphi2full}
\Omega_{de}(a)=\frac{a^{-3(1+w_0)}
\left(\frac{1+\alpha_0
a^\xi}{1+\alpha_0}\right)^{\frac{3w_0}{\xi}}}{B^{-1}a^{-3}+AB^{-1} g(a)+
a^{-3(1+w_0)}
\left(\frac{1+\alpha_0 a^\xi}{1+\alpha_0}\right)^{\frac{3w_0}{\xi}}},\ \
\end{eqnarray}
where again
$A=\Omega_{dm0}/\Omega_{b0}$ and $B=\Omega_{de0}/\Omega_{b0}$.
Finally, using Eq. (\ref{deceleration00}) the deceleration parameter is
written as
\begin{eqnarray}\label{deceleration2full}
q=-1+\frac{3}{2}\left[\frac{a^{-3}+A
g(a)+B(1+w_0)a^{-3(1+w_0)}
\left(\frac{1+\alpha_0
a^\xi}{1+\alpha_0}\right)^{\frac{3w_0}{\xi}}}{a^{-3}+A
g(a)+Ba^{-3(1+w_0)}
\left(\frac{1+\alpha_0
a^\xi}{1+\alpha_0}\right)^{\frac{3w_0}{\xi}}}\right].
\end{eqnarray}

For the special case $\xi=0$, we get the energy densities of dark sectors,
\begin{equation}
\label{dedens2}
\rho_{de}=\rho_{de0}a^{-3[(1+\alpha_0+w_0)/(1+\alpha_0)]},
\end{equation}
\begin{equation}
\label{dmdens2}
\rho_{dm}=\rho_{dm0}a^{-3}\left[1+\alpha_0\frac{\Omega_{de0}}{\Omega_{dm0}}\left(a^{-3w_0/(1+\alpha_0)}-1\right)\right].
\end{equation}
To keep $\rho_{dm}>0$ when $a\rightarrow 0$ and $a\rightarrow \infty$, the coupling constant $\alpha_0$ must satisfy the
condition $0\le \alpha_0<\Omega_{dm0}/\Omega_{de0}$.  Using Eqs. (\ref{dedens2}) and (\ref{dmdens2}), we get
\begin{equation}
\label{decqeq2}
q=\frac{1}{2}+\frac{3}{2}\frac{w_0\Omega_{de0}}{(1+\alpha_0)\Omega_{de0}+[1-(1+\alpha_0)\Omega_{de0}]a^{3w_0/(1+\alpha_0)}}.
\end{equation}
For the special case $\xi=0$, we find that when $w_0\Omega_{de0}<-1/3$,
and $\alpha_0>-3w_0-1>0$ or $\alpha_0<-1$, the cosmic acceleration is transient.
Therefore, the transient acceleration happens when the parameters satisfy the conditions:
(1) $\alpha_0> 0$ if $\xi>0$; (2) $\Omega_{dm0}/\Omega_{de0}>\alpha_0>-3w_0-1>0$ if $\xi=0$.

\begin{figure*}[htp]

\centerline{\includegraphics[width=0.8\textwidth]{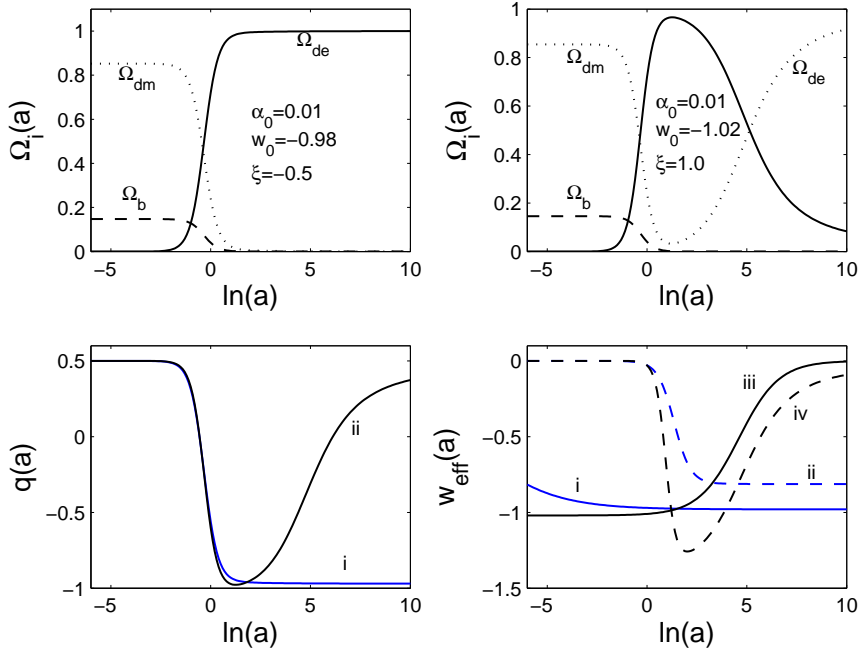}}

\caption{The results for the general interacting model $Q= \alpha_{0} a^{\xi}(\dot{\rho}_{de}+3H \rho_{de})$.
Upper left panel: The evolutions of the various density parameters
for $\alpha_0=0.01$, $\xi=-0.5$ and $w_0=-0.98$.
Upper right panel: The evolutions of the various density parameters
for $\alpha_0=0.01$, $\xi=1.0$ and $w_0=-1.02$.
Lower left panel: The corresponding
evolutions of the deceleration parameter $q$. Line (i) is for the parameters
$\alpha_0=0.01$, $\xi=-0.5$ and $w_0=-0.98$ and line (ii) is for the parameters
$\alpha_0=0.01$, $\xi=1.0$ and $w_0=-1.02$. Lower right panel: the evolutions
of the effective equation of state parameters for dark energy (lines (i) and (iii)) and dark matter (lines (ii) and (iv)).
Lines (i) and (ii) are for the parameters
$\alpha_0=0.01$, $\xi=-0.5$ and $w_0=-0.98$, and lines (iii) and (iv) are for the parameters
$\alpha_0=0.01$, $\xi=1.0$ and $w_0=-1.02$.
}
\label{mod3}
\end{figure*}

Since we obtain analytical expressions for the
evolutions of the various density parameters and of the
deceleration parameter, we proceed to present the corresponding evolutions to
show the validity of the above analysis by choosing some particular values of the parameters as examples.
In the upper left panel of Fig.~\ref{mod3} we plot the evolutions of the
density parameters with $\alpha_0=0.01$, $w_0=-0.98$ and $\xi=-0.5$, corresponding to the energy transfer from
dark energy to dark matter with decreasing transfer rate.
In the lower left panel of Fig.~\ref{mod3}, we depict the
corresponding evolution of the deceleration parameter (line (i)). As we can see, the
late-time cosmic acceleration is permanent.

In the upper right panel of Fig.~\ref{mod3}, we plot the evolutions of the
density parameters with $\alpha_0=0.01$, $w_0=-1.02$ and $\xi=1.0$.
As expected, the energy transfer from dark energy to dark matter
with increasing rate leads to dark matter domination in the future.
Furthermore, in the lower left panel of  Fig.~\ref{mod3}
we present the evolution of the deceleration parameter (line (ii)). From
these plots we can clearly see that the current acceleration of the
universe is transient for $\xi>0$. As we discussed above, the transient acceleration happens for $\xi>0$
and $\alpha_0 > 0$.
For the special case $\xi=0$, the transient acceleration happens if $\Omega_{dm0}/\Omega_{de0}>\alpha_0>-3w_0-1>0$.

In the lower right panel of Fig.~\ref{mod3}, we show the evolutions of $w_{eff}$
for both dark energy and dark matter. We see that $w^{eff}_{de}$ increases to zero
and $w^{eff}_{dm}$ decreases to be negative first and then increases back to zero again
due to the energy transfer from dark energy to dark matter when transient acceleration happens.
In the dark energy domination case, both $w^{eff}_{de}$ and $w^{eff}_{dm}$ are negative.

\section{Conclusions}
\label{Conclusions}

In the present work we investigated cosmological scenarios in which
dark matter and dark energy interact with each other by time-dependent
interaction forms, and we obtained analytical
expressions for the evolutions of the deceleration and the various density
parameters. The resulting cosmological behavior proves to be
very interesting.

In the case of a simple time-dependent interaction of the form  $Q=3\beta_0
a^\xi H \rho_{de}$, for negative $\beta_0$, the energy transfer from
dark matter to dark energy leads to a dark energy dominated universe,
independent of the values of $\xi$ and $w_0$ (of course $w_0$ should
satisfy the condition $w_0\Omega_{de0}<-1/3$ so that current cosmic acceleration can be explained,
and $\xi < 3w_0$ in order to keep $\rho_{dm}$ positive).
In this case the late-time cosmic acceleration is permanent.
However, for positive $\beta_0$ and $\xi$, the energy transfer from dark energy to dark matter
leads to a late-time dark-matter domination. In this case the current cosmic acceleration
presents a transient character, alleviating the coincidence problem.
For the special case $\xi=0$, the transient acceleration happens when $-w_0\Omega_{dm0}/(\Omega_{dm0}+\Omega_{de0})>\beta_0>-w_0-1/3>0$.
However, the observational constraints on $\Omega_{de0}$, $\Omega_{dm0}$ and $w_0$ tell us that $-w_0\Omega_{dm0}/(\Omega_{dm0}+\Omega_{de0})<-w_0-1/3$.
So we find that the transient acceleration happens if the parameters $\xi$ and $\beta_0$ are in the following ranges:
$\xi>0$ and $\beta_0>0$.

In the case of a more general interaction form
$Q= 3\alpha_{0} a^{\xi}H \rho_{de}+ \alpha_{0} a^{\xi} \dot{\rho}_{de}$,
to keep $\rho_{de}$ real, we require $\alpha_0>0$ when $\xi\neq 0$.
We find that for negative $\xi$, the
universe is led to a complete dark energy domination
with a permanent late-time acceleration. On the other hand, for
the case $\xi>0$ and $\alpha_0>0$, the energy transfer from dark energy to dark matter
leads to a dark matter domination in the far future, with a transient
cosmic acceleration. For the special case $\xi=0$, the transient acceleration happens
for $\Omega_{dm0}/\Omega_{de0}>\alpha_0>-3w_0-1>0$.
However, the observational constraints on $\Omega_{de0}$, $\Omega_{dm0}$ and $w_0$ tell us that $\Omega_{dm0}/\Omega_{de0}<-3w_0-1$.
Therefore, the transient acceleration happens if the parameters satisfy the conditions:
$\alpha_0> 0$ and $\xi>0$.

In the case of late time permanent acceleration, $w^{eff}_{dm}$ can either be positive or negative,
but $w^{eff}_{de}$ is always negative.
However, in the case of a transient acceleration, $w^{eff}_{dm}$ decreases to be negative and then
increases to zero, and $w^{eff}_{de}$ increases to be non-negative, so both $w^{eff}_{dm}$
and $w^{eff}_{de}$ are close to be non-negative in the far future.

The transient character of the cosmic acceleration is very interesting
from both the phenomenological and observational point of view, and moreover it
can offer an explanation for the recent indications that the current
cosmic acceleration is slowing down.
The time-dependent interaction with the specific phenomenological forms
examined above, offers an alternative way of obtaining the
transient acceleration comparing to other mechanisms proposed in the
literature.

\begin{acknowledgments}
This work was partially supported by the National Basic Science Program (Project 973) of
China under grant No. 2010CB833004, the NNSF project of China under
grant Nos. 10935013, 11005165 and 11175270, the Program for New Century Excellent Talents in University,
the Fundamental Research Funds for the Central Universities, and CQ CMEC under grant No. KJ110523.
\end{acknowledgments}


\end{document}